\documentclass[aps,10pt,prl,notitlepage,tightenlines,nofootinbib,twocolumn,superscriptaddress]{revtex4-2}

\usepackage{amsmath}
\usepackage{amssymb,amsthm}
\usepackage{mathtools}
\usepackage{mathrsfs}
\usepackage{relsize}
\usepackage{bm}

\usepackage{epsfig,graphics,graphicx,color,xcolor}
\usepackage{soul} 
\usepackage{cancel} 
\usepackage{slashed}	
\usepackage{subfigure}
\usepackage{array}
\usepackage{ragged2e}
\usepackage{lineno}
\usepackage{natbib}
\usepackage{hyperref}
\hypersetup{colorlinks=true,linkcolor=red,anchorcolor=red,citecolor=orange, filecolor=brown,urlcolor=red,bookmarksnumbered=true,
pdfview=FitB
}
\graphicspath{ {image/} }
\usepackage{enumitem}

\colorlet{darkgreen}{green!50!black}
\colorlet{brightyellow}{yellow!75!red}
\colorlet{orange}{red!50!yellow}
\colorlet{darkgray}{gray!50!black}

\def\dd{{\mathrm{d}}}

\newcommand{\half}[1][1] {\mathsmaller{\frac{#1}{2}}}

\makeatletter
\newcommand*{\transpose}{%
  {\mathpalette\@transpose{}}%
}
\newcommand*{\@transpose}[2]{%
  \raisebox{\depth}{$\m@th#1\intercal$}%
}
\makeatother 



\begin{document}

\title{Hadronic energy-momentum tensor revisited}

\author{Yang~Li}
\affiliation{Department of Modern Physics, University of Science and Technology of China, Hefei 230026, China}
%
\author{Qun Wang}
\affiliation{Department of Modern Physics, University of Science and Technology of China, Hefei 230026, China}

\author{James P. Vary}
\affiliation{Department of Physics and Astronomy, Iowa State University, Ames, Iowa 50011}

\date{\today}

\begin{abstract}
Everything gravitates and they do so through the energy-momentum tensor (EMT). However, there is no consensus on how to define the EMT of a hadron, e.g.~the proton, the fundamental building blocks of the visible world. In this work, we show that the hadronic EMT can be cast into the form of relativistic continua with spin. The fluid-like form allows a clear identification of the proper hadronic energy density, pressure and shear. A spin tensor contribution is also identified without the reference to the total angular momentum operator, in alignment with the recent development in relativistic spin hydrodynamics. The hadronic EMT is related to the quantum expectation value of the EMT operator through the convolution with the hadronic wavepacket. To factorize out the hadronic EMT, different multipole expansion schemes are explored. We show that the popular Breit-frame densities and light-front densities emerge as the monopole momentum densities from different factorization schemes. 
\end{abstract}

 \maketitle

\paragraph{Introduction}
There is currently an intense debate concerning the proper definition of the hadronic energy-momentum tensor (EMT) $t^{\alpha\beta}$, a physical observable of fundamental importance \cite{Burkardt:2002hr, Miller:2007uy, Miller:2009sg, Miller:2009qu, Miller:2010nz, Leader:2013jra, Lorce:2017xzd, Lorce:2018zpf, Lorce:2018egm, Polyakov:2018zvc, Miller:2018ybm, Lorce:2020onh, Jaffe:2020ebz, Panteleeva:2021iip, Freese:2021czn, Freese:2021qtb, Freese:2021mzg, Lorce:2021xku, Ji:2021mfb, Epelbaum:2022fjc, Panteleeva:2022khw, Li:2022hyf, Chen:2022smg, Freese:2022fat, Panteleeva:2022uii, Carlson:2022eps, Chen:2023dxp, Freese:2023jcp, Panteleeva:2023evj, Freese:2023abr, Lorce:2024ipy, Hagiwara:2024wqz}. The core of the problem is how to relate experimental measurements to the internal energy and stress distributions inside hadrons, e.g. the proton \cite{Burkert:2018bqq, Kumericki:2019ddg, Dutrieux:2021nlz, Duran:2022xag}. From a pure microscopic viewpoint, this problem concerns the proper interpretation of the experimental data \cite{Burkert:2023wzr}. On the other hand, $t^{\alpha\beta}$ dictates how these subatomic particles gravitate on a macroscopic level. Even though a direct measurement of the gravitational coupling of the proton is extremely challenging, how the energy and stress are distributed inside it has major physical consequences. For example, several definitions produce a singular energy density at the center of the pion based on perturbative QCD \cite{Miller:2009qu}, which is particularly concerning for the density defined with a sharply localized state \cite{Panteleeva:2021iip, Panteleeva:2022khw, Epelbaum:2022fjc, Panteleeva:2022uii, Panteleeva:2023evj}. Some approaches predict the change of the internal densities beyond the naïve Lorentz transformation \cite{Lorce:2020onh, Lorce:2024ipy}, which may play a role for investigating cold nuclear matter, ranging from the atomic nuclei to neutron stars. 
Furthermore, Polyakov et al. conjectured a mechanical stability condition $D(0) < 0$ for hadrons based on the stress density interpretation of the gravitational form factor  (GFF) $D(q^2)$ \cite{Polyakov:2018zvc}. However, there is no consensus on the exact physical interpretation of the GFFs obtained from Lorentz decomposition of the hadronic EMT \cite{Lorce:2017xzd, Lorce:2021xku, Ji:2021mtz, Ji:2021mfb, Ji:2022exr, Czarnecki:2023yqd}.

In this work, we revisit the definition of the hadronic EMT. Starting from the quantum expectation value (QEV) of the Belinfante-Rosenfeld EMT operator $T^{\alpha\beta}$ \cite{Freese:2021jqs}, we introduce a fluid-like hadronic EMT (\ref{eqn:hadronic_EMT}),
 which enables us to identify the physical hadronic densities, including the hadronic energy density $\mathcal E(x)$, pressure $\mathcal P(x)$ and shear $\varPi^{\alpha\beta}(x)$. For the proton, a hadron with spin, inspired by the recent development in spin hydrodynamics \cite{Montenegro:2017rbu,Florkowski:2017ruc,Hattori:2019lfp,Becattini:2018duy,Speranza:2020ilk,Fukushima:2020ucl}, we show that there also exists the contribution from the spin tensor $\mathcal S^{\alpha\beta}(x)$, which differs from the orbital angular momentum density considered in the literature \cite{Lorce:2017wkb, Polyakov:2018zvc, Leader:2013jra}.  

In the literature, one of the well adopted definitions is based on hadronic matrix elements (HMEs). 
Inspired by the well-known work of Sachs \cite{Ernst:1960zza, Sachs:1962zzc}, Polyakov et. al. defines the hadronic EMT as the Fourier transform of the HMEs in the Breit frame where $\vec p' = -\vec p = 2\vec q$  \cite{Polyakov:2018zvc}: 
\begin{equation}\label{eqn:BF_HMEs_FT}
t^{\alpha\beta}_\textsc{bf}(\vec r) = \int \frac{\dd^3 q}{(2\pi)^32E_q} e^{-i\vec q\cdot \vec r} \langle -\half \vec q| T^{\alpha\beta}(0) | +\half \vec q\rangle \,.
\end{equation}
Here, $2E_q = 2\sqrt{M^2+\frac{1}{4}\vec q^2}$ accounts for the normalization of the state vectors. 
 This definition of hadronic EMT relies on several arbitrary choices: the choice of a special reference frame and the choice of the phases of the state vectors. The latter suggests that this is not a proper density in quantum theory. Note that the proton is not at rest in this frame. Jaffe argued that this definition is only valid for nonrelativistic systems \cite{Jaffe:2020ebz}. 
 
The frame dependence can be eliminated by a projection onto the light front $t-z/c$ \cite{Miller:2007uy, Miller:2009sg, Miller:2009qu, Miller:2010nz,  Miller:2018ybm,  Freese:2021czn, Freese:2021qtb, Freese:2021mzg, Freese:2022fat, Freese:2023jcp, Freese:2023abr}, 
\begin{equation}\label{eqn:LF_projections}
T^{\alpha\beta}_\textsc{lf}(\vec x_\perp) = \int \dd x^- \, T^{\alpha\beta}(x^-, \vec x_\perp)
\end{equation}
where, $x^\pm = x^0 \pm x^3$, $\vec x_\perp = (x^1, x^2)$ are the light-front coordinates.  The light-front hadronic EMT is defined as the 2D Fourier transform of the light-front-projected operator, 
\begin{multline}\label{eqn:LF_HMEs_FT}
t^{\alpha\beta}_\textsc{lf}(\vec r_\perp) = \frac{1}{2P^+} \int \frac{\dd^2 q_\perp}{(2\pi)^2} e^{-i\vec q_\perp\cdot \vec r_\perp} \\
\times \langle P -\half \vec q| T^{\alpha\beta}_\textsc{lf}(0) |P +\half \vec q\rangle \,.
\end{multline}
Here $2P^+$ accounts for state normalization. Hadronic wave functions evolving in light front time $x^+$ are boost invariant. As a result, the $P$-dependence of the light-front hadronic EMT factorizes. For example, the light-front energy density of the proton can be written as \cite{Freese:2022fat, Cao:2023ohj, Xu:2024}, 
\begin{multline}\label{eqn:LF_t+-}
t^{+-}_\textsc{lf}(\vec r_\perp) = \\
\frac{\vec P^2_\perp \mathcal A(r_\perp) + \vec P_\perp \cdot \big[\nabla_\perp\times\vec {\mathcal S}(r_\perp)\big]+ \mathcal M^2(r_\perp)}{P^+}\,,
\end{multline}
where, $\mathcal A(r_\perp)$, $\vec{\mathcal S}(r_\perp)$ and $\mathcal M^2(r_\perp)$ are intrinsic densities independent of the average momentum $P$. Comparing to the light-front energy of a free hadron $p^- = (\vec p^2_\perp + M^2)/p^+$, $\mathcal A(r_\perp)$ and $\mathcal M^2(r_\perp)$ can be interpreted as the number density and invariant mass squared density, respectively. The $\vec{\mathcal S}$ term is a surface term, representing the spin distribution within the proton. 

The above definition can be further improved by adopting a fully covariant decomposition. The HMEs of the EMT operator for the proton can be written as \cite{Kobzarev:1962wt, Pagels:1966zza, Polyakov:2018zvc}, 
\begin{multline}\label{eqn:EMT_HMEs}
\langle p', s'|T^{\mu\nu}(0)|p, s\rangle = 
\frac{1}{2M}\overline u_{s'}(p') \Big[
2P^\mu P^{\nu} A(q^2)  \\
+ iP^{\{\mu}\sigma^{\nu\}\rho}q_\rho J(q^2) 
+ \frac{1}{2} (q^\mu q^\nu - q^2 g^{\mu\nu}) D(q^2) \Big] u_s(p)
\end{multline} 
where, $a^{\{\mu}b^{\nu\}} = a^\mu b^\nu + a^\nu b^\mu$, $P = (p'+p)/2$, $q = p'-p$. 
The Lorentz scalars $A(q^2)$, $J(q^2)$ and $D(q^2)$ are known as GFFs. 
The light-front densities and the Breit frame densities can be expressed in terms of the GFFs. 
It is tempting to interpret the Fourier transforms of GFFs $A(q^2)$, $J(q^2)$ and $D(q^2)$ as the mass, spin and pressure distributions of the proton, respectively. However, this interpretation suffers from several problems. 

First, the GFFs depend on $q^2$ the four-momentum transfer squared, whereas the densities are defined with spatial dimensions. One solution is to go back to the Breit frame in which $q^0 = 0$, and $q^2 = -\vec q^2$ only involves the spatial momentum $\vec q$. However, this choice is not physically motivated. Indeed, any frame with a fixed average momentum $\vec P = (\vec p + \vec p')/2$ also fixes $q^0$, and the Breit frame gives the most compact expression. 
Alternatively, one can get back to the projection onto the light front, where $q^2 = -q^2_\perp$ only involves the 2D transverse momentum $\vec q_\perp = (q_x, q_y)$. 

Second, HMEs are not quantum expectation values (QEVs) and the GFFs resulting from the covariant decomposition of HMEs are not physical observables. Indeed, the covariant decomposition is not unique and the physical interpretation of the associated GFFs is ambiguous. Note that these ambiguities cannot be resolved by resorting to the HMEs,~e.g.~Eqs.~(\ref{eqn:BF_HMEs_FT} and \ref{eqn:LF_HMEs_FT}), since they are not Lorentz invariants. 

The situation is in stark contrast to the EMT of continua, e.g. solids and fluids, \cite{Rezzolla:2013dea}
\begin{multline}\label{eqn:continua_EMT}
t^{\alpha\beta} = e u^\alpha u^\beta  - p \Delta^{\alpha\beta} + \half \partial_\sigma ( u^{\{\alpha} s^{\beta\}\sigma} ) + \pi^{\alpha\beta} \\
+ \, \text{dissipative terms}.
\end{multline}
Here, $u^\alpha$ is the four-velocity of the continuum ($u_\mu u^\mu = 1$), and $\Delta^{\alpha\beta} = g^{\alpha\beta} - u^{\alpha}u^{\beta}$ is the spatial metric tensor. Lorentz scalar $e$ is the proper energy density, viz., the energy density measured in the local rest frame, and $p$ is the proper pressure. In ideal hydrodynamics, only the first two terms are present. The traceless shear tensor $\pi^{\alpha\beta}$ encodes the dissipative effect in fluids, but it may exist in solids as a non-dissipative quantity. For fluids with spin, it was recently proposed that the spin degree of freedom also contributes to the EMT through the spin tensor $s^{\alpha\beta}$ \cite{Fukushima:2020ucl}. 
Eq.~(\ref{eqn:continua_EMT}) is applicable to general classical continuum subject to Poincaré symmetry. 
In contrast to $t^{\alpha\beta}$, the proper densities $e$, $p$, $s^{\alpha\beta}$ and $\pi^{\alpha\beta}$ are physical densities that characterize the internal structure of the continuum independent of its external motion. 
We will show below that the hadronic EMT can also be rewritten in the form of Eq.~(\ref{eqn:continua_EMT}). As such, these physical densities can be identified within hadrons. 

\paragraph{Hadronic energy-momentum tensor} When gravity interacts with quantum matter, the metric tensor $g^{\alpha\beta}$ couples the QEV of the EMT operator. In particular, for the nucleon, the QEV is,
\begin{equation}\label{eqn:HEV}
t^{\alpha\beta}(x) = \langle\Psi|T^{\alpha\beta}(x)|\Psi\rangle,
\end{equation}
where $|\Psi\rangle$ is a normalized hadronic eigenstate of the QCD Hamiltonian. Using the covariant decomposition of the HME (\ref{eqn:EMT_HMEs}) and the covariant formalism introduced in our previous work~\cite{Li:2022hyf}, the QEV (\ref{eqn:HEV}) can be written as $t^{\alpha\beta}(x) = \big\langle \mathcal T^{\alpha\beta}(x) \big\rangle_\Psi$, 
where $\langle \cdots \rangle_\Psi$ represents a convolution with respect to the hadronic wavepacket $\Psi(x)$, 
\begin{equation}\label{eqn:convolution}
\big\langle \mathcal O(x) \big\rangle_\Psi = \int \dd^3z\, \overline\Psi(z)\mathcal O(x-z) \Psi(z)\Big|_{x^0=z^0}\,,
\end{equation}
and the kernel 
\begin{equation}\label{eqn:hadronic_EMT}
\mathcal T^{\alpha\beta} = \mathcal E \mathcal U^\alpha \mathcal U^\beta - \mathcal P \Delta^{\alpha\beta}  + \half \partial_\rho \big( \mathcal U^{\{\alpha} \mathcal S^{\beta\}\rho} \big)
+  \varPi^{\alpha\beta} 
\end{equation} 
formally represents the pure hadronic contribution, and will be referred to as the \emph{hadronic energy-momentum tensor} in a similar form of the EMT of the continuum (\ref{eqn:continua_EMT}). Therefore, $\mathcal E, \mathcal P, \mathcal S^{\alpha\beta}, \varPi^{\alpha\beta}$ can be uniquely identified as the physical densities,~i.e.~the energy, pressure, spin and shear densities, respectively. 
The dissipative terms are not relevant for hadrons. 
For the nucleon, the hadronic wavepacket $\Psi(x)$, 
\begin{equation}\label{eqn:hadronic_wavepacket}
\Psi(x) = \sum_s \int \frac{\dd^3p}{(2\pi)^32p^0} e^{-ip\cdot x} u_s(p) \langle p, s | \Psi\rangle
\end{equation} 
satisfies the Dirac equation. 

The convolution (\ref{eqn:convolution}) highlights the wave nature of hadrons. Jaffe argued that the Compton wavelength of the proton $\lambda_\text{Comp} = M^{-1}_p \simeq 0.2\,\text{fm}$ is comparable to their matter radii $r_p \simeq 0.5\,\text{fm}$  \cite{Jaffe:2020ebz}.
Consequently, under a sub-femtometer probe $\lambda_\gamma \ll r_p \sim \lambda_\text{Comp} \ll \lambda_\text{deBrog}$,  the proton behaves as a wave. This is in contrast to nonrelativistic systems, e.g. nuclei, where the de Broglie wavelength $\lambda_\text{deBrog}$ can be taken to be much smaller than the wavelength of the probing photon $\lambda_\gamma$, and the system behaves as a structured particle. 

The hadronic 4-velocity $\mathcal U^\alpha$ is defined from the conserved number current $n^\mu = (1/2M)\overline\Psi i\tensor\partial^\mu \Psi = \langle n \, \mathcal U^\mu \rangle_\Psi$ with normalization $\mathcal U_\alpha \mathcal U^\alpha = 1$, where $f \tensor\partial g = f\partial g - (\partial f) g$ is the convective derivative, and $n$ is the proper number density. 
The hadronic densities can be obtained from the usual projection technique. For example, the total energy density $e(x) = \langle \mathcal U_\alpha \mathcal U_\beta \mathcal T^{\alpha\beta} \rangle_\Psi \equiv  \langle \mathcal E(x) \rangle_\Psi$.  
Plugging Eqs.~(\ref{eqn:EMT_HMEs}) and (\ref{eqn:hadronic_wavepacket}) into 
Eq.~(\ref{eqn:HEV}), we identify the physical hadronic densities as,  
\begin{align}
\mathcal E(x) =\,& M\int\frac{\dd^3q}{(2\pi)^3} e^{iq\cdot x} \Big\{\Big(1 - \frac{q^2}{4M^2}\Big)A(q^2) \nonumber \\ 
& + \frac{q^2}{4M^2}\Big[2J(q^2) - D(q^2)\Big]\Big\}, \label{eqn:energy_density}\\
\mathcal P(x) =\,& \frac{1}{6M}\int\frac{\dd^3q}{(2\pi)^3} e^{iq\cdot x} q^2 D(q^2), \label{eqn:pressure}\\
\mathcal S^{\alpha\beta}(x) =\,& \int\frac{\dd^3q}{(2\pi)^3} e^{iq\cdot x} \nonumber \\
& \times \Big\{
i\sigma^{\alpha\beta} \sqrt{1 - \frac{q^2}{4M^2}}  
  -  \frac{U^{[\alpha} q^{\beta]}}{2M} \Big\} J(q^2), \label{eqn:spin_density} \\
\mathcal \varPi^{\alpha\beta}(x) =\,& \frac{1}{4M}\int\frac{\dd^3q}{(2\pi)^3} e^{iq\cdot x} \Big(q^\alpha q^\beta - \frac{q^2}{3} \Delta^{\alpha\beta} \Big) D(q^2)\,, \label{eqn:stress_density}
\end{align}
where $U^\alpha = i\tensor\partial^\alpha/\sqrt{4M^2-q^2}$, $\Delta^{\alpha\beta} = g^{\alpha\beta} - U^\alpha U^\beta$, and $a^{[\mu}b^{\nu]} = a^\mu b^\nu - a^\nu b^\mu$. %
 Note that both the hadronic spin tensor $\mathcal S^{\alpha\beta}$ and shear tensor $\varPi^{\alpha\beta}$ explicitly depend on the convective derivative $\tensor\partial$, hence, on the hadronic wavepacket $\Psi(x)$. 

Following the values of the GFFs in the forward limit, i.e., $A(0) = 1, J(0) = 1/2$, the energy density $\mathcal E(x)$ is normalized to the hadron mass $M$, and the integral of the pressure density $\mathcal P(x)$ vanishes \cite{Laue:1911lrk, Lorce:2017xzd}. 
The energy density can be used in the energy conditions to impose a constraint on $D$ \cite{Lorce:2018egm}, although it is known that energy conditions are violated in the quantum realm. From these densities, we can introduce the trace scalar density $\theta = g_{\alpha\beta}t^{\alpha\beta}$, which satisfies the standard relation with the energy density and pressure,
$\theta = \big\langle \varTheta(x) \big\rangle =  \big\langle \mathcal E(x)   -  3 \mathcal P(x) \big\rangle$.

The hadronic spin tensor $\mathcal S^{\alpha\beta}$ satisfies the Frenkel condition $\langle \mathcal U_\alpha \mathcal S^{\alpha\beta} \rangle_\Psi = 0$ \cite{Frenkel:1926zz}.  It consists of two terms: a term $\propto \sigma^{\alpha\beta}$ representing the intrinsic spin contribution and a term $\sim \partial^{[\alpha} \mathcal U^{\beta]}$ representing the vorticity contribution.  Similar terms appear in the medium polarization tensor of electromagnetic current but with different form factors, viz. they are associated  with different spatial distributions \cite{ Li:2022hyf}.  The spatial part of the spin tensor $\mathcal S^{ij}(r)$ is normalized to the familiar spin vector $\frac{i}{2}\sigma^{ij} \sim \frac{1}{2}\epsilon^{ijk}\mathcal S^k$ for on-shell Dirac spinors, which induces the spin current $\nabla \times \vec{\mathcal S}$ shown in (\ref{eqn:LF_t+-}). 
 
\paragraph{Hadronic multipole moment densities} The hadronic densities in (\ref{eqn:hadronic_EMT}) are in convolution with the hadronic wavepacket. To extract these densities from the physically measurable QEV, the factorization of the intrinsic densities and the wavepacket is required. For example, such factorization exists in nonrelativistic dynamics, e.g. nuclei, and one can uniquely define the internal nuclear distribution by deconvoluting the c.m.~wavepackets from the QEV \cite{Cockrell:2012vd}. 
On the other hand,  the hadronic densities Eqs.~(\ref{eqn:energy_density}--\ref{eqn:stress_density}) depend on the fluid velocity $\mathcal U^\alpha$ through 
\begin{equation}\label{eqn:q0_P}
q^0 = \sqrt{(\vec P + \half \vec q)^2+M^2} - \sqrt{(\vec P - \half \vec q)^2+M^2},
\end{equation}
where $\vec P = ({i}/{2})\tensor\nabla \propto \vec{\mathcal U}$.  Consequently, the hadronic densities and hadronic wavepacket do not factorize. 

To reinstate factorization, several approaches are proposed. Lorcé showed that factorization exists in the transverse phase space \cite{Lorce:2020onh}. 
Alternatively, factorization also exists on the light front, thanks to the Galilean nature of the light-front boosts \cite{Freese:2021czn, Freese:2021qtb, Freese:2021mzg, Freese:2022fat, Freese:2023jcp, Freese:2023abr}. 

Within the covariant approach, we proposed the asymptotic factorization 
in the spirit of the QCD factorization of hard and soft physics. 
We perform a multipole expansion in terms of the convective derivative $i\tensor\nabla$:
\begin{equation*}\label{eqn:multipole_expansion}
\mathcal O(x, i\tensor\nabla) = \mathcal O_0(x) + \mathcal O_1^{i}(x) i\tensor\nabla^i + \frac{1}{2!}\mathcal O_2^{ij}(x) i\tensor\nabla^i i\tensor\nabla^j + \cdots
\end{equation*}
The expansion is based on the separation of the intrinsic hadronic scale $l_\text{had} = \max\{r_p, \lambda_\text{Comp}\}$ from the scale (width) of the wavepacket $l_\Psi = \lambda_\text{deBrog}$. 
Since 
\begin{equation}\label{eqn:q0_BF}
q_0 \sim O\Big(\frac{l_\text{had}^2}{l_\Psi^2}\Big), \; q^2 = -\vec q^2 + O\Big(\frac{l_\text{had}^2}{l_\Psi^2}\Big)\,,
\end{equation} 
the leading term of the expansion -- the monopole term factorizes out, i.e. it is independent of the wavepacket.  The extracted monopole energy density is, 
\begin{multline} \label{eqn:energy_density_Sachs}
\mathcal E_0(\vec r) = M\int\frac{\dd^3q}{(2\pi)^3} e^{-i\vec q\cdot \vec r} \Big\{\Big(1 + \frac{\vec q^2}{4M^2}\Big)A(-\vec q^2)  \\ 
 - \frac{\vec q^2}{4M^2}\Big[2J(-\vec q^2) - D(-\vec q^2)\Big]\Big\}\,,
\end{multline}
where the subscript ``0" indicates that these are the monopole densities. 
Note that now the 4-momentum squared $q^2$ is replaced by $-\vec q^2$ the 3-momentum squared. The above monopole energy density agrees with the Breit-frame energy density $t^{00}_\textsc{bf}(r_\perp)$.  
Similarly, the monopole hadronic spin, pressure and shear densities also agree with the corresponding Breit-frame densities \cite{Polyakov:2018zvc}. 
Therefore, the Breit-frame hadronic densities can be properly interpreted as the monopole moment densities of the  QEVs of the EMT operator \cite{Li:2022hyf}. 
Note that the above derivation goes beyond the non-relativistic approximation. 
 Higher multipole densities also exist, and are expected to play a role in cases where the factorization of the monopole moment is not sufficient, such as nucleons within the deuteron or in neutron stars. 

The factorization discussed above is not unique. 
Another useful choice is to expand around a highly boosted wavepacket with momentum $|\vec P_0| \gg l^{-1}_\text{had}$.  
In this case, the 4-momentum squared $q^2$ becomes, 
\begin{equation}
q^2 = -q^2_\perp + O\Big(\frac{l^{-2}_\text{had}}{|\vec P_0|^2}, \frac{l^2_\text{had}}{l^2_\Psi}\Big), 
\end{equation}
where $\vec q_\perp = \vec q - (\vec q\cdot \hat n)\hat n$ and $\hat n = \vec P_0/|\vec P_0|$. 
It is advantageous to adopt the light-front coordinates for a 4-vector $v^\mu$, 
\begin{equation}
v^\pm = v^0 \pm v_\|, \quad \vec v_\perp = \vec v - v_\|\hat n\,,
\end{equation}
where, $v_\| = \hat n \cdot \vec v$. 
For $\hat n = \hat z$, these variables become the standard light-front variables. 
The leading-order hadronic energy density becomes two-dimensional, 
\begin{multline}\label{eqn:energy_density_IMF} 
\mathcal E_0(x) = \delta(x_\|) M \int \frac{\dd^2 q_\perp}{(2\pi)^2}  e^{-i \vec q_\perp \cdot \vec x_\perp}  \Big[\Big(1 + \frac{\vec q^2_\perp}{4M^2}\Big) A(-\vec q^2_\perp)  \\
 - \frac{\vec q^2_\perp}{4M^2}\Big(2J(-\vec q^2_\perp) - D(-\vec q^2_\perp)\Big)\Big]. 
 \end{multline}
which agrees with the light-front energy density $t^{+-}_\textsc{lf}(r_\perp)$, reaffirming the equivalence between densities in the infinite momentum frame (IMF) and densities on the light front.  

In principle, one can adopt a wavepacket with a finite central momentum $\vec P_0$, which is applicable to, e.g., fast moving nucleons in a nucleus. 
The 4-momentum squared $q^2$ becomes, 
\begin{equation}
q^2 = \big(\hat q^0\big)^2 - \vec q^2 + O\Big(\frac{l^2_\text{had}}{l^2_\Psi}\Big)
\end{equation}
where, $\hat q^0$ is be obtained from $q^0$ by replacing $\vec{P}$ with $\vec{P}_0$ in (\ref{eqn:q0_P}). The Breit-frame and light-front/IMF densities can be viewed as hadronic densities within this boosted frame for two limiting cases $|\vec P_0| \to 0$ and $|\vec P_0| \to \infty$, respectively \cite{Lorce:2020onh}. 

\paragraph{Summary and discussions}

The hadronic energy-momentum tensor (EMT) is of fundamental importance in modern physics. A proper definition of this quantity is obscured by the relativistic wave nature of hadrons. In this work, starting from the quantum expectation value (QEV) of the Belinfante-Rosenfeld EMT operator, we proposed a covariant definition of the hadronic EMT. The resulting expression resembles the continua, e.g. fluids and solids, which enables us to unambiguously define the physical hadronic densities, such as the proper energy density,  pressure, spin density and shear. Our fluidlike EMT also supports the use of the coupled multifluid picture for decomposing the proton mass into different contributions \cite{Rezzolla:2013dea, Lorce:2017xzd, Ji:2021mtz, Ji:2021qgo, Lorce:2021xku}. 

To factorize out the hadronic EMT from the hadronic QEVs, we also adopt multipole expansions. The energy and stress densities introduced in the non-covariant approach within the Breit frame is recovered as the monopole momentum densities. Similarly, the boost invariant light-front densities are recovered as the monopole moment densities from the infinite-momentum limit. 

The hadronic spin tensor $\mathcal S^{\alpha\beta}$ identified here is new and is consistent with the recent progress in relativistic spin hydrodynamics. 
One should distinguish this tensor density   
 from the generalized orbital angular momentum (OAM) density $\ell^{\lambda,\alpha\beta} = \langle \Psi| L^{\lambda, \mu\nu} |\Psi\rangle$. The generalized angular momentum $J^{\lambda, \mu\nu} = L^{\lambda, \mu\nu} + S^{\lambda, \mu\nu}$, the conserved Noether current associated with the Lorenz symmetry, can be decomposed into an orbital part $L^{\lambda, \mu\nu} = x^\mu T^{\lambda \nu} - x^\nu T^{\lambda \mu}$ and a spin part $S^{\lambda, \mu\nu}$. In the Belinfante form, the OAM density coincides with the total angular momentum density. Therefore, $\ell^{\lambda,\alpha\beta}$ is widely adopted to investigate the spin composition of hadrons (see Ref.~\cite{Leader:2013jra} for a review of this topic). 
 The spin tensor $s^{\mu\nu}$ originates from the spin part of the generalized angular momentum $s^{\lambda,\mu\nu} = \langle\Psi| S^{\lambda, \mu\nu} |\Psi\rangle = u^\lambda s^{\mu\nu}$ \cite{Jeon:2023mlv}.  The spin tensor contributes to the EMT as a spin current: $t^{0i}_\text{spin} \sim (\nabla \times \vec{\mathcal S})^i$ \cite{Fukushima:2020ucl}. 

Our definition of the hadronic EMT and hadronic densities can also be applied to hadrons with a different spin. For (pseudo-)scalar hadrons, e.g. the pion, our covariant formalism again gives the same result for the hadronic EMT (\ref{eqn:hadronic_EMT}), except there is no spin term. The physical hadronic densities are also identical to Eqs.~(\ref{eqn:energy_density}, \ref{eqn:pressure}, \ref{eqn:stress_density}) with $J(q^2) = 0$. One should be mindful, however, these densities differ from those introduced by Polyakov et al. in a non-covariant approach for spin-0 hadrons due to different normalization conventions \cite{Polyakov:2018zvc}. 

\paragraph*{Acknowledgements}

The authors acknowledge fruitful discussions with Adam Freese, Gerald Miller, Cédric Lorcé, Kirill Tuchin, Chandan Mondal, Siqi Xu and Xianghui Cao. Y. L. is supported by the New faculty start-up fund of the University of Science and Technology of China. This work was supported in part by the National Natural Science Foundation of China (NSFC) under Grant No. 12375081 and 12135011, by the Chinese Academy of Sciences under Grant No. YSBR-101, and by the U.S. Department of Energy (DOE) under Grant No. DE-SC0023692.

\end{document}